\documentclass[aps,prl,reprint,groupedaddress,showpacs,showkeys,floatfix]{revtex4-1}
\usepackage{watermark}
 \usepackage{amsmath}
 \usepackage{amsfonts}
 \usepackage{graphicx}
 \usepackage{epsfig}
 \usepackage{epstopdf}
 \usepackage{comment}
 \usepackage{newtxtext,newtxmath}

 \usepackage{hyperref}
 \hypersetup{
    bookmarks=true,         
    unicode=false,          
    pdftoolbar=true,        
    pdfmenubar=true,        
    pdffitwindow=false,     
    pdfstartview={FitH},    
    pdftitle={Probing positron cooling in noble gases via annihilation gamma spectra},   
    pdfauthor={D. G. Green},     
    pdfsubject={PosGamCool},   
    pdfcreator={D. G. Green},   
    pdfproducer={Producer}, 
    pdfkeywords={keywords}, 
    pdfnewwindow=true,      
    colorlinks=true,       
    linkcolor=blue,          
    citecolor=blue,        
    filecolor=magenta,      
    urlcolor=blue           
}

\begin{document}
\title{Probing positron cooling in noble gases via annihilation $\gamma$ spectra}
\author{D.~G. Green}
\email[]{d.green@qub.ac.uk}
\affiliation{
Centre for Theoretical Atomic, Molecular and Optical Physics,
School of Mathematics and Physics,\\
Queen's University Belfast, Belfast BT7\,1NN, Northern Ireland, United Kingdom}
\date{\today}

\begin{abstract}
$\gamma$ spectra for positron annihilation in noble-gas atoms are calculated using many-body theory for positron momenta up to the positronium-formation threshold. 
This data is used, together with time-evolving positron-momentum distributions determined in [arXiv:1706.01434v1 (2017)], to calculate the time-varying $\gamma$ spectra produced during positron cooling in noble gases. 
The $\gamma$-spectra and their $\bar{S}$ and $\bar{W}$ shape parameters are shown to be sensitive probes of the time evolution of the positron momentum distribution, and thus provide a means of studying positron cooling that is complementary to positron lifetime spectroscopy.
\end{abstract}

\maketitle

Low-energy positrons annihilate with atomic electrons forming two $\gamma$ rays whose Doppler-broadened energy spectra are characteristic of the electron state involved, e.g., annihilation on tightly-bound core electrons contributes to the high-Doppler-shift wings of the spectrum \cite{DGG:2015:core}. 
This gives positrons important use in e.g., non-destructive high-sensitivity studies of surfaces, defects and porosity of industrially important materials \cite{RevModPhys.60.701,RevModPhys.66.841,RMPpossolids2013,hugreview}.
Importantly, however, the $\gamma$ spectra are also characteristic
of the positron momentum at the instant of annihilation: increased positron momentum results in an increased annihilating-pair momentum, leading to larger $\gamma$-ray Doppler shifts.
Measurement of the time-varying $\gamma$ spectra, or so called AMOC (`Age MOmentum Correlation')  spectra \cite{amoc,posbeams,AMOC:1997,Engbrecht:AMOC,AMOC:2016}), in which both the positron `age' (i.e., lifetime from source to annihilation) and $\gamma$ spectra are measured in coincidence, can thus enable the study of positron and positronium cooling in atomic gases (see e.g., \cite{AMOC:psmt,AMOC:ar}).
Understanding the dynamics of positron cooling in gases is critical for the accurate interpretation of experiments, and for the development of efficient positron cooling in traps and accumulators \cite{AlQaradawi:2000}, and a cryogenically cooled, ultra-high-energy-resolution, trap-based positron beam \cite{Natisin:2014,Natisin:2016}. 

Positron cooling in atomic gases has traditionally been probed by positron annihilation lifetime spectroscopy (PALS) (see \cite{Griffith:1979, Charlton:1985} for reviews), which involves measurement of the spectrum of positron lifetimes. 
The dynamics of positron cooling in noble gases was recently elucidated in \cite{DGG_cool}, where Monte-Carlo (MC) simulations based on accurate scattering and annihilation cross sections calculated using many-body theory (MBT) were used to determine the time-evolving positron momentum distribution and normalized annihilation rate $\bar{Z}_{\rm eff}(t)$. That work found that a strikingly small fraction of initial positrons survive to thermalization, affecting the measured positron annihilation rate, and explaining the discrepancy between trap-based and gas-cell measurements in Xe. Overall, good agreement was found with the long-standing PALS measurements for all the atoms except Ne, for which the calculated cooling time was found to be drastically longer than the measured value. It was proffered that the discrepancy was due to an incorrect analysis of the experimental data and/or the presence of impurities. 
New experiments are called for to further test the theoretical results. 
Verifying the accuracy of the calculations is important to ensure that the complicated positron-atom many-body system is well understood. 

In this letter we use MBT to investigate the dependence of the positron annihilation $\gamma$ spectra for the noble-gas atoms on the positron momentum, up to the positronium (Ps) formation threshold, and demonstrate that the time-varying $\gamma$ spectra provide a sensitive probe of positron cooling in noble gases that is complementary to PALS. 
The MBT takes full account of positron-atom and positron-electron correlations, including virtual-positronium formation \cite{PhysScripta.46.248,dzuba_mbt_noblegas,DGG_posnobles,DGG:2015:core}. 
Specifically, we extend the calculations of \cite{DGG:2015:core}, where $\gamma$ spectra for thermal positron annihilation in individual core and valence subshells of the noble gases were calculated using MBT. It provided an accurate description of the measured spectra for Ar, Kr and Xe and firmly established the relative contributions of various atomic orbitals to the spectra. 
Using the spectra calculated at all positron momenta, together with the time-evolving positron-momentum distributions calculated using MBT based MC simulations in \cite{DGG_cool}, we calculate the $\gamma$ spectra produced during positron cooling. 
We analyze the dynamics of the the $\bar{S}$ and $\bar{W}$ shape parameters, which characterize the low and high (two-$\gamma$) momentum parts of the spectra, during the process of positron thermalization.
The present results provide benchmarks to which positron-cooling experiments can compare. 

\emph{Many-body theory calculations of annihilation $\gamma$ spectra.---}In the dominant process, a positron of momentum ${\bf k}$ and energy $\varepsilon=k^2/2$ annihilates with an electron in state $n$ to form two $\gamma$-ray photons of total momentum ${\bf P}$ \cite{qed}. In the center-of-mass frame the two $\gamma$ rays have equal energies $mc^2=511$ keV (neglecting the initial positron and electron energies). In the laboratory frame the photon energies are Doppler shifted by $\epsilon \leq Pc/2$, and their spectrum is \cite{Dunlop:gamma,DGG:2015:core}
\begin{eqnarray}\label{eqn:w}
w_{nk}(\epsilon) = 
\frac{1}{c}\int_{2|\epsilon|/c}^{\infty}
\int_{\Omega_{\bf P}} |A_{n\varepsilon}({\bf P})|^2\frac{d\Omega_{\bf P}}{(2\pi)^3}PdP,
\end{eqnarray}
where $A_{n\varepsilon}({\bf P})$ is the annihilation amplitude. 
It is calculated via a diagrammatic expansion [see Fig.~(1) of \cite{DGG:2015:core}], 
including the zeroth-order vertex, and the first- and higher-order (``$\Gamma$-block") corrections, which account for the attractive electron-positron interaction at short range \cite{Dunlop:gamma,DGG:2015:core,DGG_corelong}. 
The total spectrum is given by the sum over the holes $w_{k}(\epsilon) = \sum_n w_{nk}(\epsilon)$. Its integral gives the effective annihilation rate $Z_{\rm eff}(k)=\int_{-\infty}^{\infty}w_k(\epsilon) d\epsilon$ \cite{Fraser,Pomeranchuk,Dunlop:gamma,DGG:2015:core}

\begin{figure}[t!!]
\begin{center}
\includegraphics[width=0.38\textwidth]{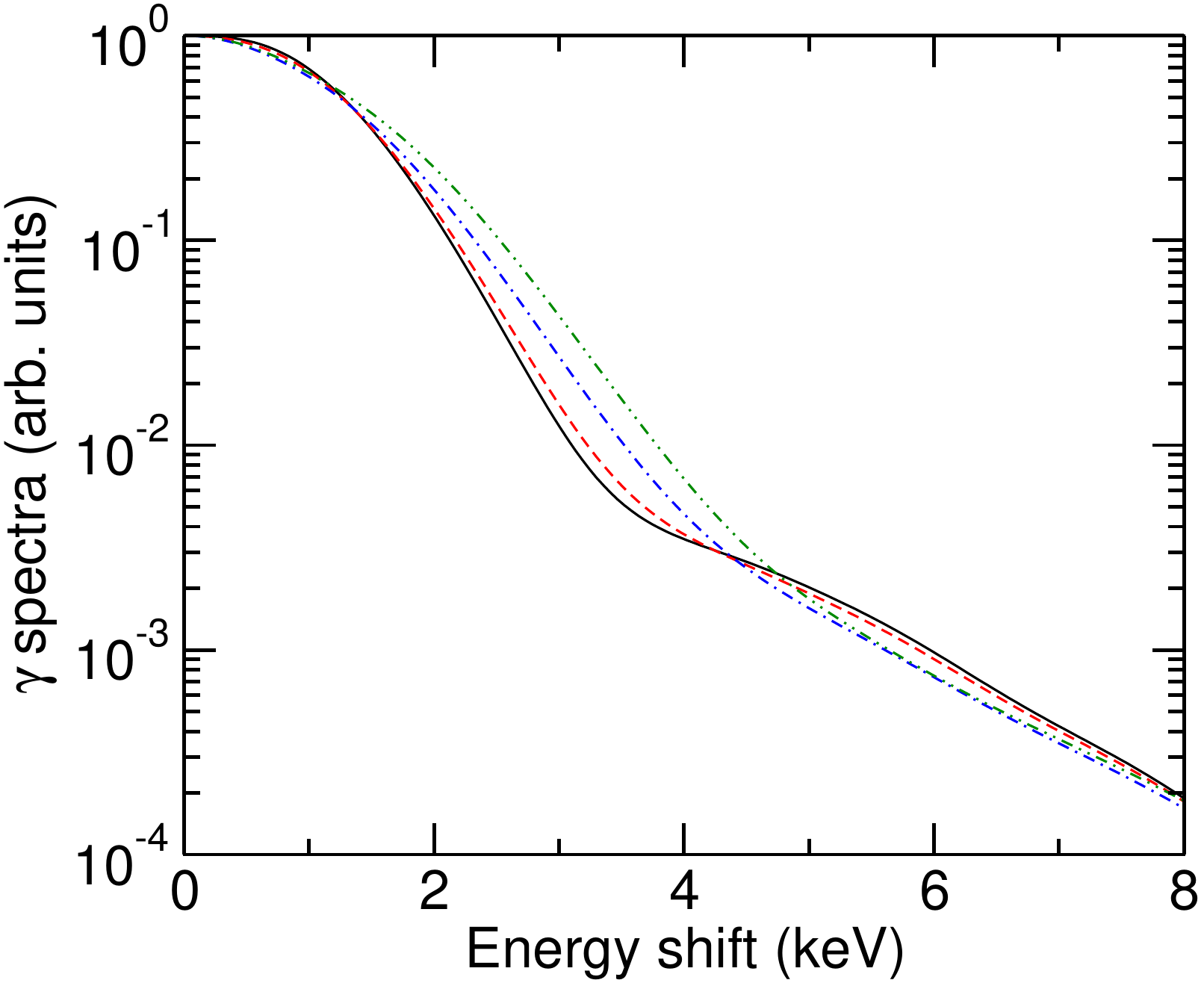}\\
\caption{
$\gamma$ spectra for positron annihilation in Ar, for positron of momentum $k=0.04$ a.u. (black-solid line), $k=0.2$ a.u. (red-dashed line), $k=0.4$ a.u. (blue-dashed-dotted line) and $k=0.6$ a.u. (green dash-dash-dotted line). 
\label{fig:arspec}}
\end{center}
\end{figure}

To illustrate the momentum dependence of the shape of the $\gamma$ spectra, Fig.~\ref{fig:arspec} shows the MBT calculated $\gamma$ spectra for argon, for $k=0.04$--$0.6$ a.u., normalized to unity at zero energy shift. 
The calculations included $s$, $p$ and $d$-wave incident positrons (higher partial waves contribute negligibly) \footnote{After integrating over the directions of ${\bf P}$, all positron partial waves contribute to the amplitude incoherently, such that it can be calculated independently for each positron angular momentum $\ell_{\varepsilon}$.} annihilating on the valence $ns$ and $np$, and subvalence $(n-1)s, (n-1)p$, and $(n-1)d$ subshells, e.g., in Ar: the 3$s$ and 3$p$ valence, and 2$s$ and 2$p$ subshells.
In general, at a given positron momentum the spectra are characteristic of the electron orbitals involved
 e.g., annihilation with core electrons produces a broader component than that with valence electrons, contributing to the distinct shoulders in the spectrum \cite{DGG:2015:core}.
The figure shows that the $\gamma$ spectrum is significantly broadened for higher positron momenta (see also Fig.~20 of \cite{DGG_hlike}). Increasing the positron momentum leads to increased momenta of the annihilating electron-positron pair, and allows the positron to penetrate deeper into the atomic core, ultimately resulting in larger Doppler shifts. 
The exact shape of the spectra is somewhat complicated. The broadening of both the core and valence contributions with the positron momentum is accompanied by the increasing relative importance of $p$ and $d$ wave contributions as $k$ increases, whose spectra are typically narrower than the $s$-wave one \cite{DGG_corelong},\footnote{At low momenta the Wigner threshold law predicts the partial rate $Z_{\rm eff}^{(\ell_{\varepsilon})}(k)\sim k^{2\ell_{\varepsilon}}$ \cite{quantummechanics} and thus the $s$-wave contribution dominates the spectra.
However, the contribution of $p$ and $d$-waves increases and become comparable at positron momenta close to the positronium-formation threshold.}. 

\begin{figure}[h!!]
\begin{center}
\includegraphics[width=0.37\textwidth]{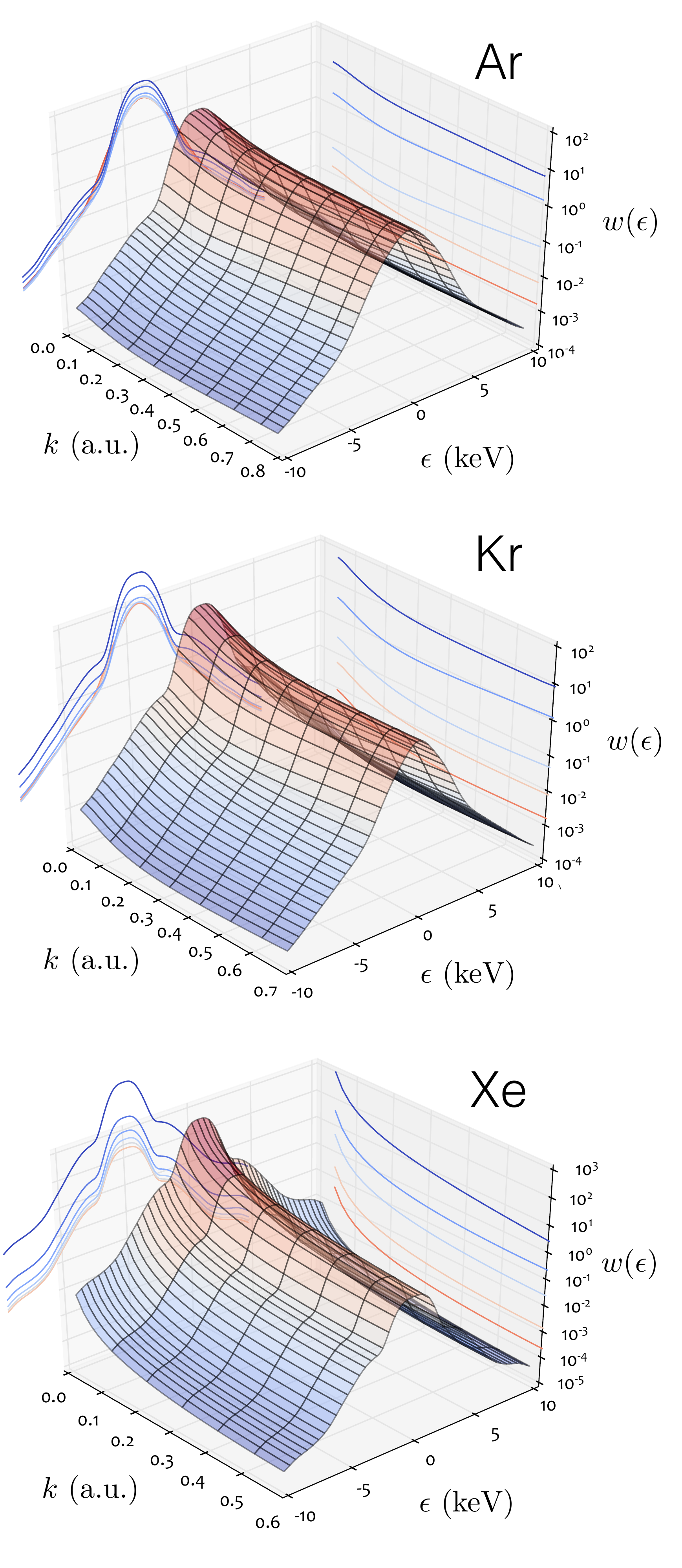}\\
\caption{Calculated $\gamma$ spectra $w_k(\epsilon)$ for positron annihilation on Ar, Kr and Xe as a function of positron momentum $k$ up to the respective positronium formation thresholds. Also shown are projections  at $\epsilon=0,2,4,6$, and 8 keV, and 
at $k=0.02$ a.u.~and $k=0.1$~a.u.~up to the Ps-formation threshold in step sizes of 0.1 a.u. 
 \label{fig:nob3d}}
\end{center}
\end{figure}

Figure \ref{fig:nob3d} shows the absolute MBT calculated $\gamma$-spectra for Ar, Kr and Xe 
\footnote{
The $\gamma$ spectra for thermal positrons presented in \cite{DGG:2015:core} were found to be in excellent agreement with experiment \cite{PhysRevLett.79.39}. A slight discrepancy was, however, observed in the wings of the spectrum for Xe, where the MBT underestimated the experimental result. It was proffered that this may be remedied by considering the thermally averaged spectrum. 
It has been calculated in the present work for all the noble gases, but however, has been found to be almost indistinguishable from the $\gamma$ spectra calculated using positrons with $k=0.04$ a.u. 
It remains to investigate whether the discrepancy is thus due to the neglect of relativistic effects that may enhance the core annihilation spectra. 
Relativistic MBT calculations are required to confirm this.}. 
The increase in magnitude of the spectra as $k\to0$ accords with the rise of the effective annihilation rate $Z_{\rm eff}(k)$ (see Fig.~2 in \cite{DGG_cool}). 
This effect is due to the existence of positron-atom virtual levels \cite{Goldanski}, signified by large scattering lengths for Ar, Kr and Xe (see Table I in \cite{DGG_posnobles}).

\begin{figure}[t!!]
\begin{center}
\includegraphics[width=0.37\textwidth]{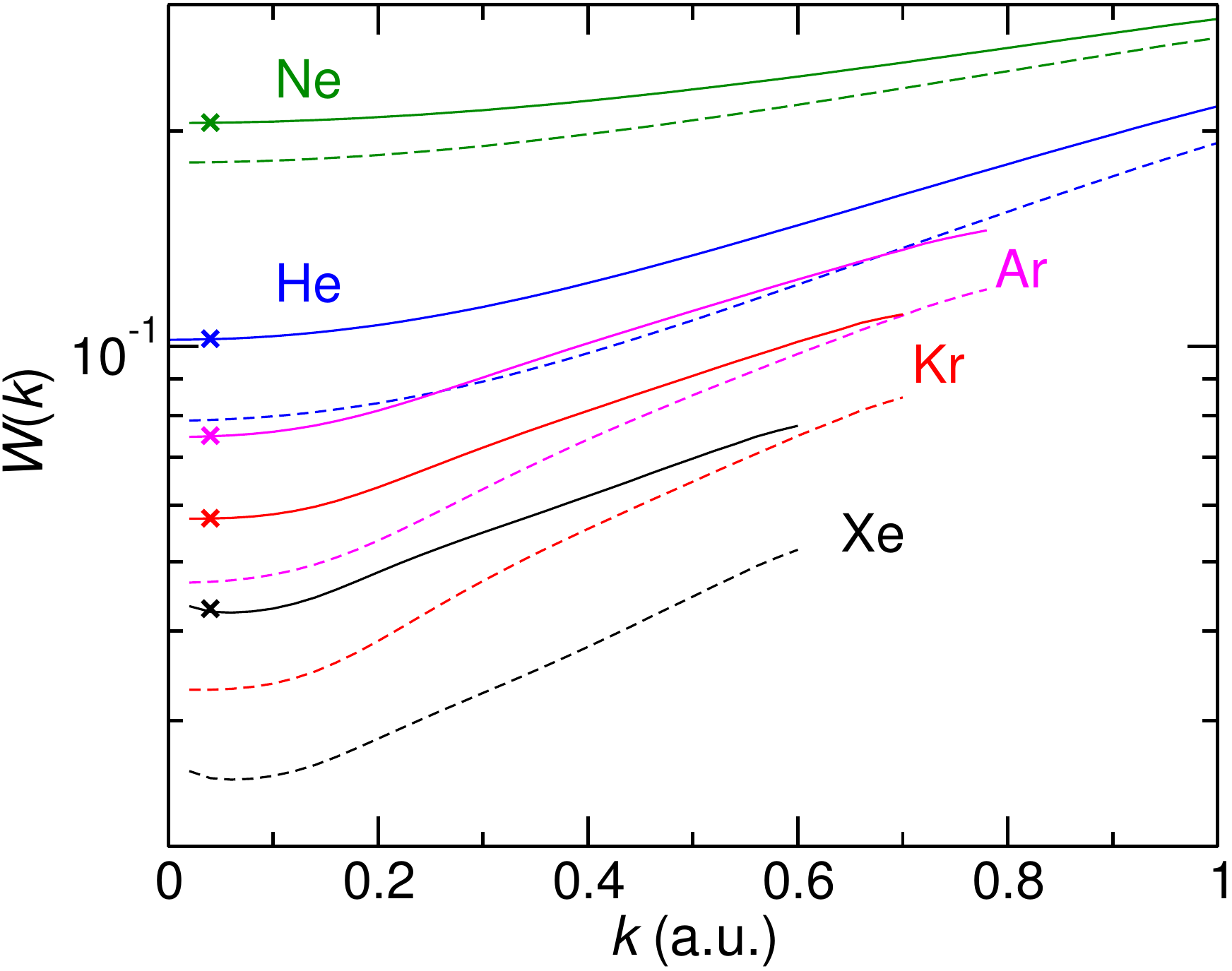}
\caption{Calculated $W(k)$ for raw and detector-convolved $\gamma$ spectra (dashed and solid lines, respectively) for positron annihilation on helium (blue); neon (green); argon (magenta); krypton (red); xenon (black). 
Also marked are the values for the thermally-averaged (at $T=293$ K) detector-convolved spectra  (crosses). \label{fig:swall}}
\end{center}
\end{figure}

The momentum dependence of the spectra can be characterized through the dimensionless parameters 
$W(k) = {2{Z_{\rm eff}(k)}^{-1}\int_{\epsilon_{W}}^{\infty} w_k(\epsilon)d\epsilon}$ and 
$S(k) = {2{Z_{\rm eff}(k)}^{-1}\int_0^{\epsilon_{S}} w_k(\epsilon)d\epsilon}$,
where $\epsilon_W$ and $\epsilon_S$ are constants. 
$W$ parameterizes the high (two-$\gamma$) momentum `wing' part of the spectrum, which originates from annihilation with core electrons and with valence electrons when they have larger momenta at smaller, core radii.
$S$ parametrizes the low two-$\gamma$-momentum region of the spectrum, which originates predominantly from annihilation on valence electrons.
Figure \ref{fig:swall} shows the calculated $W(k)$ for He--Xe for the raw $\gamma$ spectra and that convolved with the typical Ge detector-resolution function $D(\epsilon)=N\exp[-(\epsilon/a\Delta E)^2]$, where $a=1/(4\ln 2)^{1/2}$, $N=(a\Delta E\sqrt{\pi})^{-1}$ is a normalization constant, and $\Delta E$=1.16 keV \cite{PhysRevLett.79.39}, with $\epsilon_W=2.0$ keV. 
It is clear that $W(k)$ is sensitive to the positron momentum, increasing monotonically with $k$ by a factor of 1.5--2 up to the Ps-formation threshold for all the noble gases considered. 
It decreases across the sequence Ne--Xe (He is an exception since it has no core electrons). 
In contrast, the probability of annihilation on core electrons $P_{\rm core} = Z_{\rm eff}^{\rm core}/Z_{\rm eff}$, where $Z_{\rm eff}^{\rm core}$ is the annihilation rate on the core subshells, is found to increase from Ne to Xe, and decrease with positron momentum \cite{DGG_corelong}.
Thus the momentum dependence of $W(k)$ is dominated by the contribution of the positron momentum itself, rather than the change in the relative core annihilation probability.

\emph{Positron cooling probed via time-varing $\gamma$ spectra.---}
The time-varying $\gamma$ spectrum produced by positrons cooling in gases is
$\bar{w}_{\tau}(\epsilon) = {\int_0^{\infty} f(k,\tau) w_{k}(\epsilon)} dk$,
where $\tau$ is the time (typically quoted in ns) scaled by the number density of the gas (in amagat): $\tau=nt$, and $f(k,\tau)$ is the positron momentum-space distribution. It is normalized as $\int f(k,\tau)dk=F(\tau)$, the fraction of initial positrons remaining.
The momentum distributions $f(k,\tau)$ for positron cooling in noble gases were calculated recently in \cite{DGG_cool} via MC simulations based on the accurate MBT cross sections.
There it was shown that, for all the noble gases, positrons rapidly bunch around the minimum in the coefficient $B(k)= k\sigma_{\rm t}k_{\rm B}Tm/M$, where $\sigma_{\rm t}$ is the momentum-transfer cross section (see Fig.~1 of \cite{DGG_cool}), where the cooling rate slows, making the overall cooling times somewhat insensitive to the exact form of the initial distribution. After bunching in the minima, the positrons cool further slowly, before evolving towards the Maxwell-Boltzmann distribution. 
The characteristic trajectory followed by the positrons in $(k,\tau)$ space, along with the dependence of the $\gamma$ spectra on the positron momentum, leads to a characteristic AMOC spectrum i.e., the number of $\gamma$ rays $\tilde{N}_{\gamma}$ (per unit positron) detected per unit time and Doppler-shifted energy. It can be measured in experiments \cite{amoc,posbeams,AMOC:1997,Engbrecht:AMOC,AMOC:2016,AMOC:ar} and calculated as
\begin{eqnarray}\label{eqn:amoc}
\frac{d^2{\tilde{N}_{\gamma}}}{d\tau d\epsilon}= 2\pi r_0^2cF(\tau)\bar{w}_{\tau}(\epsilon).
\end{eqnarray}
Integrating over the Doppler-shifted energy $\epsilon$ gives the lifetime spectrum (normalized to one positron) $A(\tau)= d\tilde{N}_{\gamma}/2d\tau = -dF(\tau)/d\tau = \pi r_0^2cF(\tau)\int_{-\infty}^{\infty} \bar{w}_{\tau}(\epsilon)d\epsilon = \pi r_0^2c F(\tau)\bar{Z}_{\rm eff}(\tau)$ \footnote{This holds for a spectrum convolved with a detector-resolution function $D(\epsilon)$ normalized as $\int_{-\infty}^{\infty} D(\epsilon)d\epsilon =1$.} that is traditionally measured in PALS \cite{Griffith:1979,Charlton:1985}.

\begin{figure}[t!!]
\begin{center}
\includegraphics[width=0.46\textwidth]{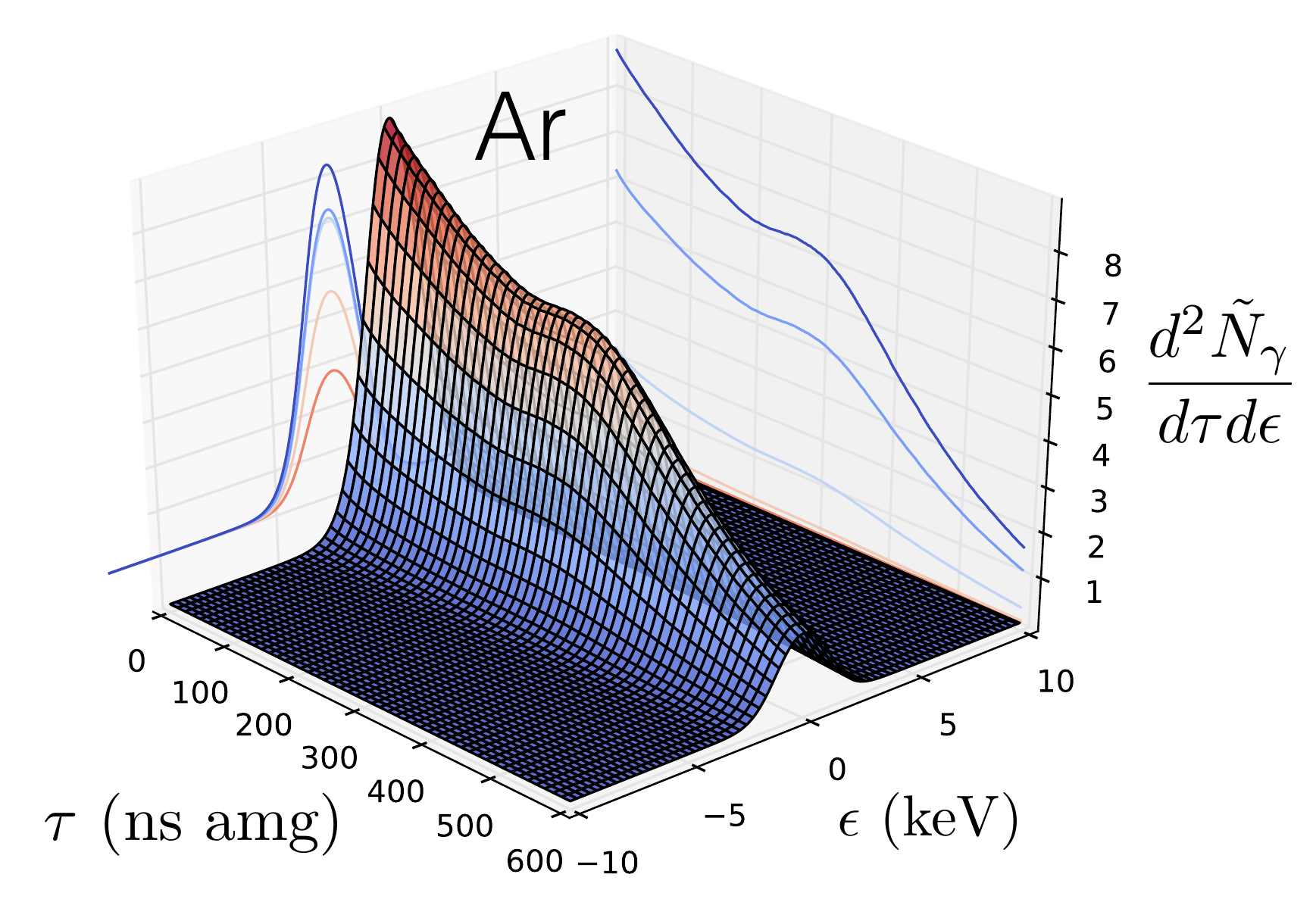}
\caption{AMOC spectrum for Ar [Eqn.~(\ref{eqn:amoc})], in units of [ns amg keV]$^{-1}$/$\pi r_0^2c$, calculated using the detector-convolved spectra $\bar{w}_{\tau}(\epsilon)$, with positrons initially distributed uniformly in energy. Also shown are projections at $\tau = 100, 200, 300, 400$ and 500 ns and at $\epsilon = 0,1,2,3,4$ and 5 keV.
\label{fig:amoc}}
\end{center}
\end{figure}

An example AMOC spectrum is shown in Fig.~\ref{fig:amoc} for Ar (c.f.~Fig.~(3) in \cite{DGG_cool}), calculated using the detector-convolved spectra for for positrons initially distributed uniformly in energy up to the Ps-formation threshold.
The overall decrease in the magnitude of the spectra in time is due to the reduction in the number of positrons surviving $F(\tau)$ (see Fig.~4 in \cite{DGG_cool}) dominating over the increase in $w_k$ as $k\to 0$. 
The flattening of the `ridge' at $\tau \lesssim 200$ ns amg occurs as the positrons are `trapped' around the momentum-transfer cross section minima and cool slowly through it.
Beyond $\tau \sim$ 200--300 ns amg epithermal annihilation occurs at momenta below the momentum-transfer cross-section minimum.
Integrating over $\epsilon$, this `knee' in the spectrum becomes the characteristic shoulder region observed in the lifetime spectrum \cite{Falk:1964,Tao:1964,Paul:1964}. It is known to be relatively insensitive to the initial positron momentum distribution \cite{Farazdel:1977,Griffith:1979}.
At later time-densities ($\tau>400$ ns amg)  the AMOC spectrum is proportional to $F(\tau)w_{\infty}$, where $w_{\infty}=\int_0^{\infty} w_k(\epsilon)  f_{\infty}(k) dk$ and 
$f_{\infty}(k)$ is the final quasi-steady-state positron-momentum distribution \footnote{For He the final distribution is the Maxwell-Boltzmann distribution. For the heavier gases, most notably Xe, the low-momentum side of the final distribution is suppressed relative to the Maxwell Boltzmann distribution.
This is a result of the the rapid depletion of low-momentum positrons due to the vigorous increase in $Z_{\rm eff}$ at low momenta.}.

From the AMOC spectrum, the time-varying $\gamma$-spectra shape parameters
$\bar{W}(\tau)\equiv  2 \bar{Z}_{\rm eff}(\tau)^{-1}\int_{\epsilon_W}^{\infty}\bar{w}_{\tau}(\epsilon) d\epsilon$ and $\bar{S}(\tau) \equiv  2 \bar{Z}_{\rm eff}(\tau)^{-1}\int_{0}^{\epsilon_S}  \bar{w}_{\tau}(\epsilon) d\epsilon$ can be determined. 
\begin{figure}[t!!]
\begin{center}
\includegraphics[width=0.35\textwidth]{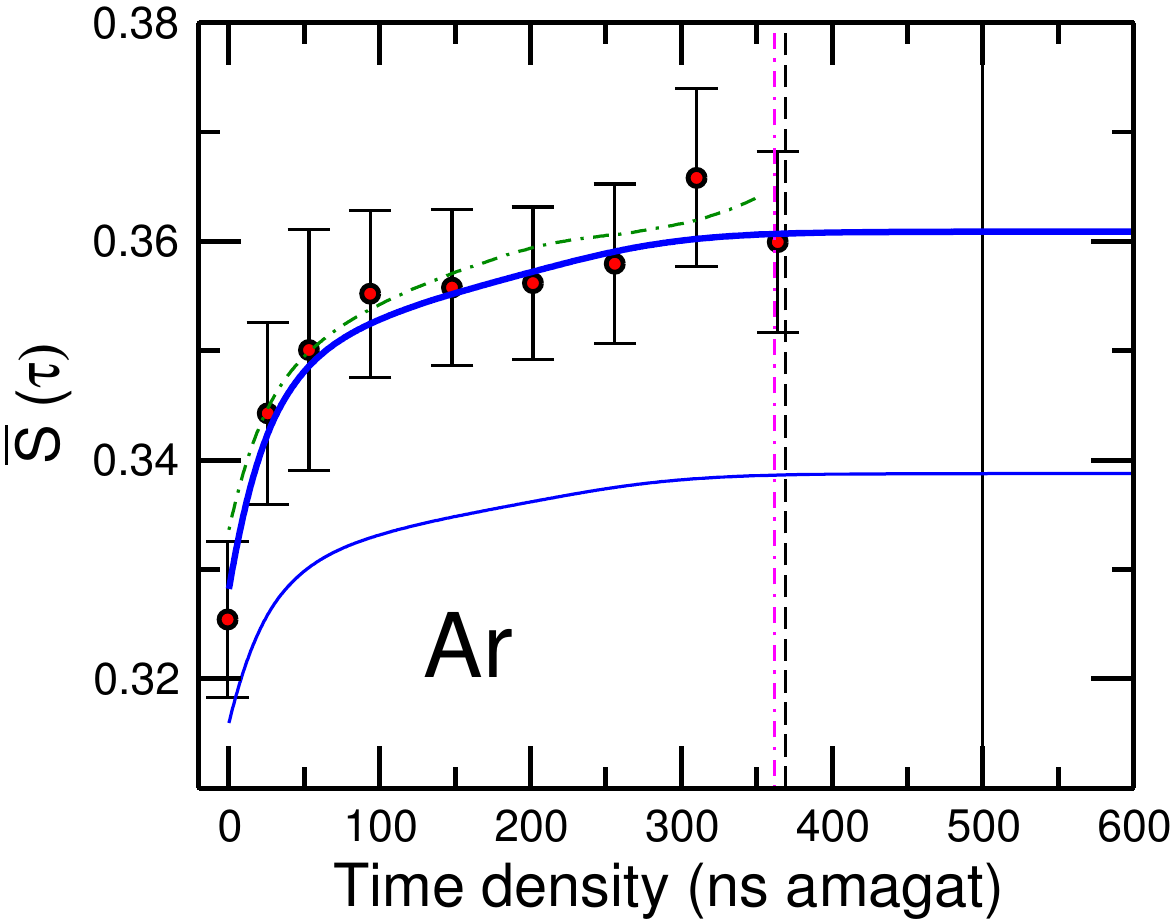}~~
\caption{$\bar{S}(\tau)$ for Ar: experiment \cite{AMOC:ar} (red circles); present calculation, for positrons distributed uniformly in energy (thin blue line); that scaled as $\bar{S}_{\rm sc} = 1.43\bar{S} - 0.13$ (thick blue solid line); calculated by first subtracting 0.1 keV$^{-1}$ from $\bar{w}_{\tau}(\epsilon)$ (green dashed-dotted line). Also marked are the calculated \cite{DGG_cool} and measured \cite{Coleman:1975} shoulder lengths (vertical dot-dashed and dashed lines), and calculated complete thermalization time (solid vertical line) \cite{DGG_cool}. \label{fig:swarexp}}
\end{center}
\end{figure}

\begin{figure}[p!!]
\begin{center}
\includegraphics[width=0.305\textwidth]{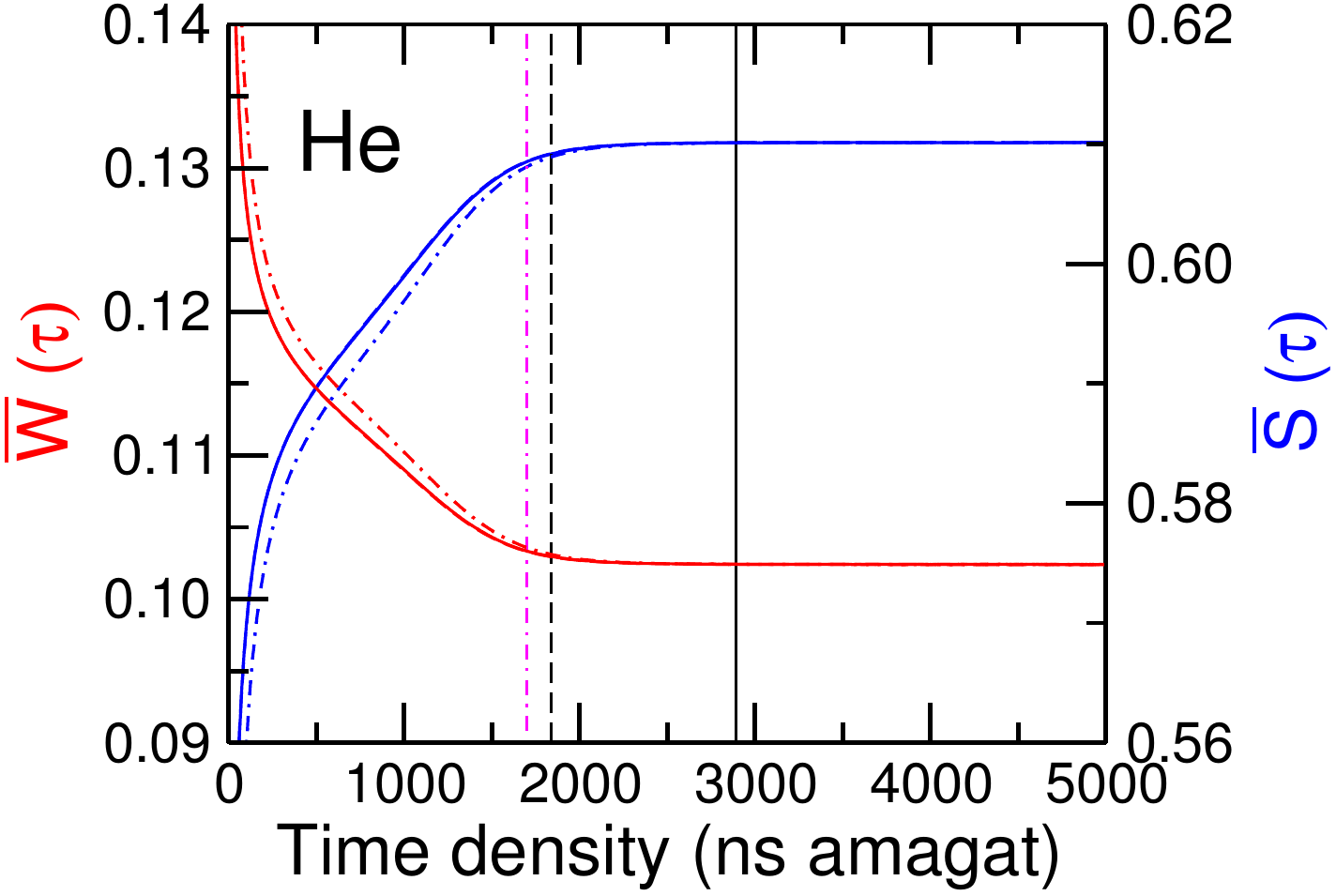}\\[0.5ex]
\includegraphics[width=0.315\textwidth]{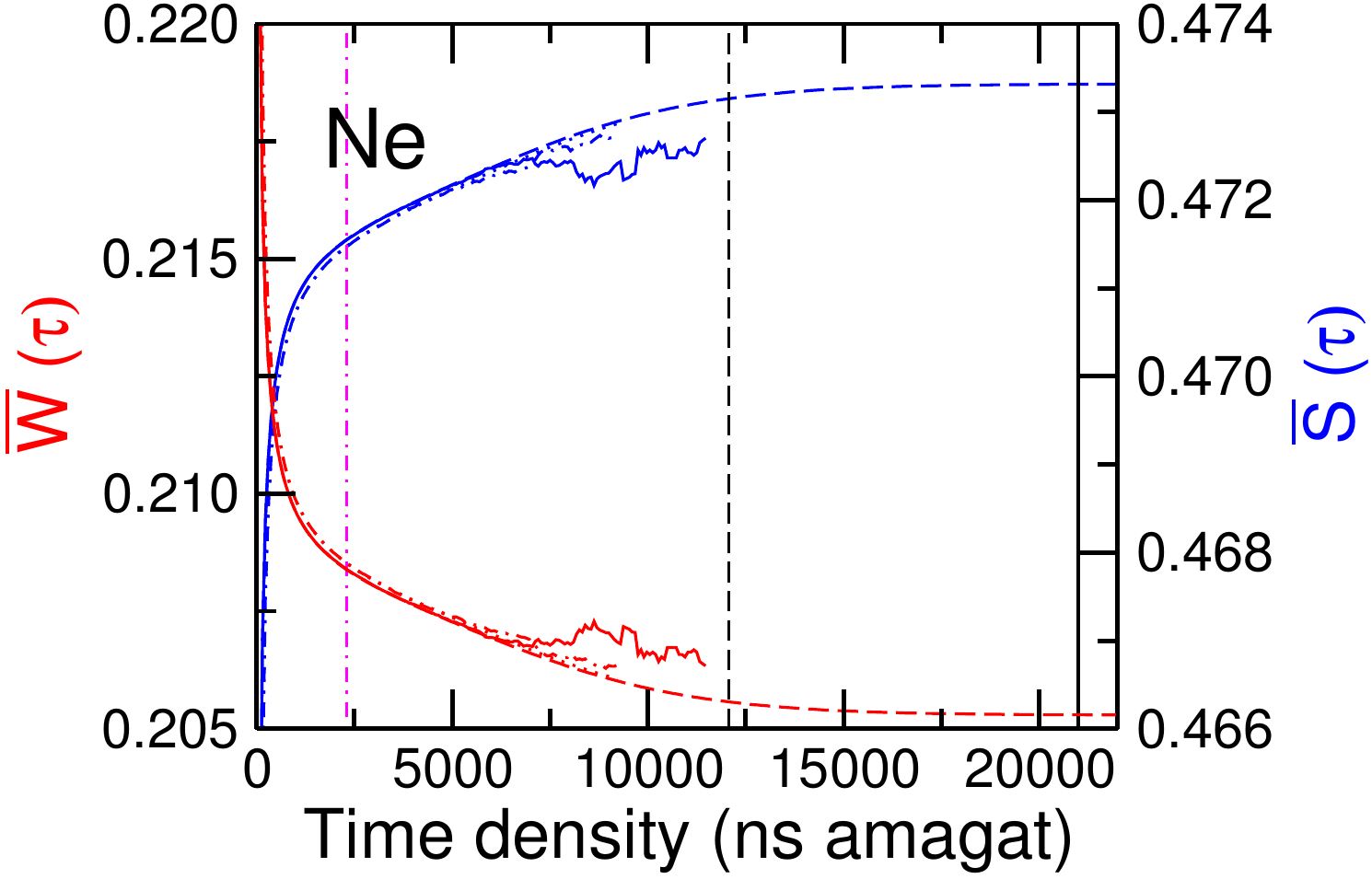}\\[0.5ex]
\includegraphics[width=0.305\textwidth]{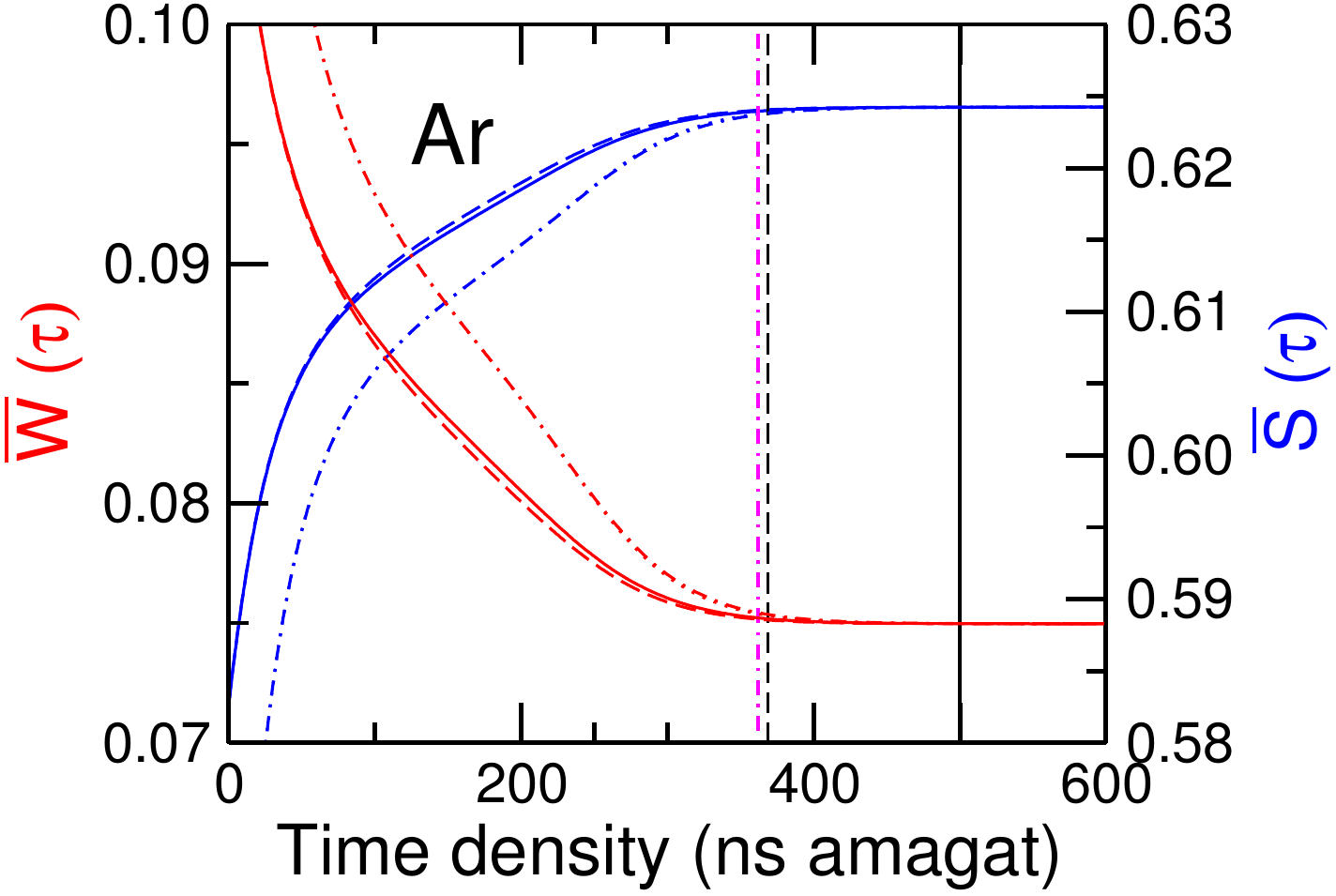}\\[0.5ex]
\includegraphics[width=0.305\textwidth]{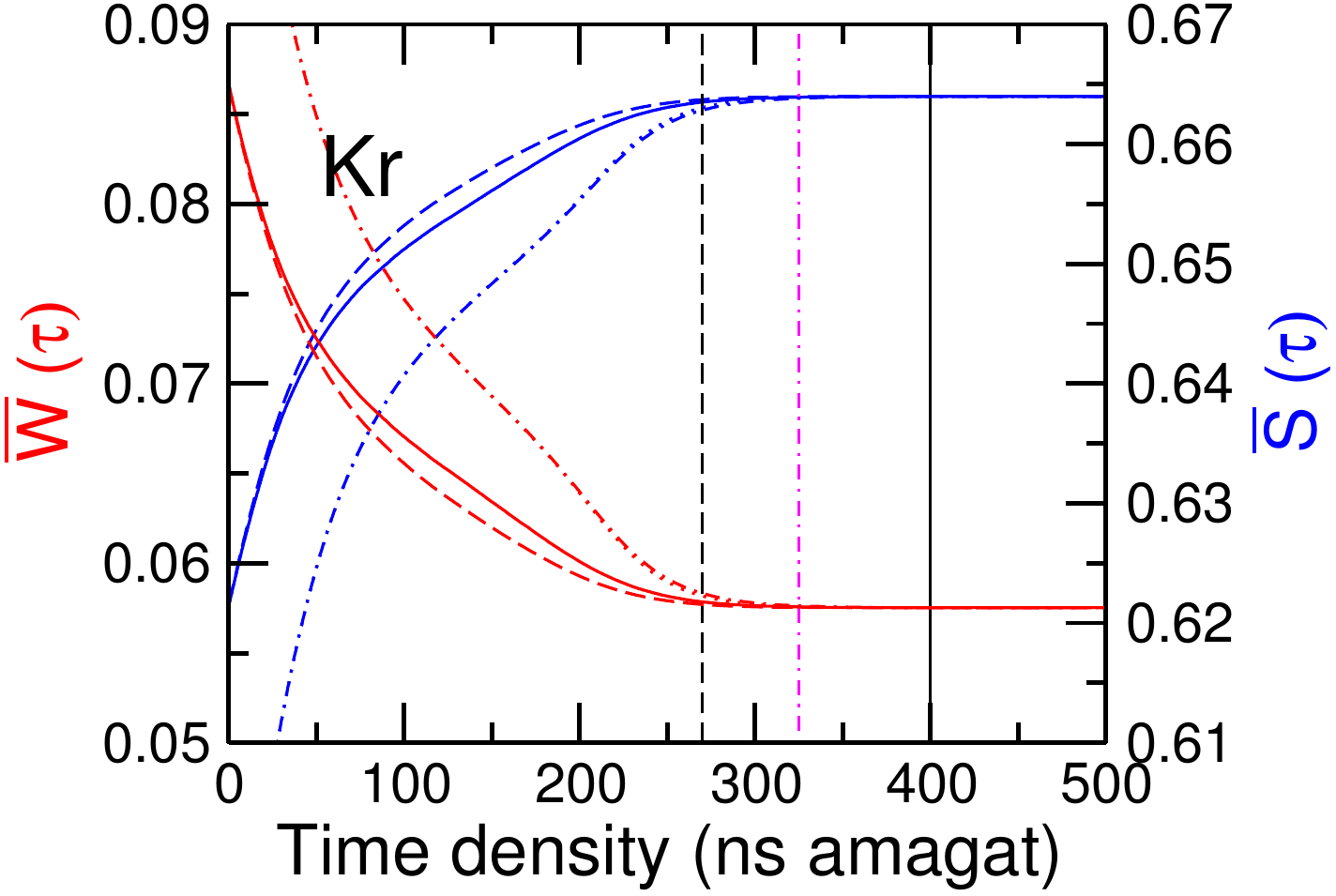}\\[0.5ex]
\includegraphics[width=0.305\textwidth]{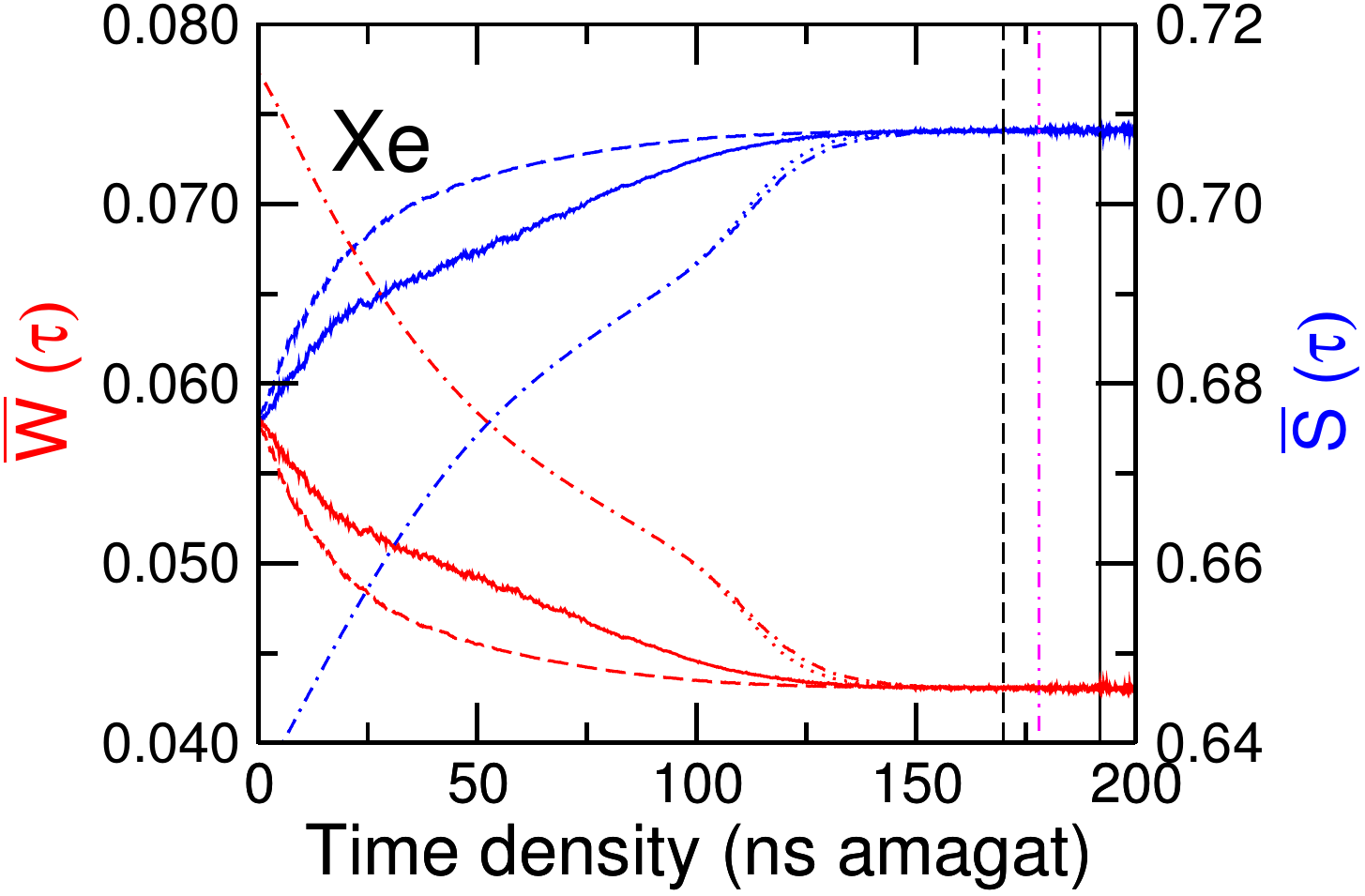}
\caption{Calculated $\bar{W}(\tau)$ and $\bar{S}(\tau)$ parameters (red and blue lines, respectively) for detector-convolved $\gamma$ spectra: excluding and including depletion of the distribution due to annihilation, for positrons initially distributed uniformly in energy (dashed and solid lines, respectively), and with energy equal to the Ps-formation threshold (dotted and dash-dotted lines, respectively, which are almost indistinguishable). 
Also shown are the shoulder lengths and thermalisation times calculated in \cite{DGG_cool} (dashed and solid vertical lines, respectively), and the experimental shoulder lengths of \cite{Coleman:1975} (He and Ne) and \cite{Wright:1985} (Ar, Kr and Xe). 
The steady-state values of $\bar{W}$ $(\bar{S})$ are for
He: 0.102 (0.610); 
Ne: 0.2067, (0.473);
Ar: 0.624, (0.075);
Kr: 0.664, (0.058);
Xe: 0.708, (0.043).
The value for Ne neglected depletion of the distribution due to annihilation. For Ne, at $\tau\sim 5000$ ns amg, where $F(\tau)\lesssim 10^{-2}$, $W(\tau)$=0.2073 and $S(\tau)$=0.472.
\label{fig:sw1}}
\end{center}
\end{figure}
Figure~\ref{fig:swarexp} shows the calculated $\bar{S}(\tau)$ compared with experiment \cite{AMOC:ar} (obtained using $\epsilon_{\rm S}=0.5$ keV). Although the theoretical value is systematically lower than the measured one, the time dependence is in near perfect agreement. This is made evident by scaling the theoretical result as $\bar{S}_{\rm sc} = 1.43\bar{S} - 0.13$. Such scaling can account for effects in background subtraction in the experiment: e.g., calculating $\bar{S}(\tau)$ by first subtracting 0.1 (ns amg keV)$^{-1}$ from the spectrum produces excellent agreement with experiment (green dashed-dotted line). 
In PALS, a common measure of the cooling time is the `shoulder length' $\tau_{\rm s}$, defined via $\bar{Z}_{\rm eff}(\tau_{\rm s}) = \bar{Z}_{\rm eff} - 0.1\Delta Z$, where $\bar{Z}_{\rm eff}$ is the final steady-state effective annihilation rate and $\Delta Z \equiv \bar{Z}_{\rm eff} - \bar{Z}_{\rm min}$, where $\bar{Z}_{\rm min}$ is the minimum of $\bar{Z}_{\rm eff}(\tau)$ \cite{Paul:1968,Griffith:1979,Charlton:1985}. 
An alternative measure is the `complete thermalization time', defined as the time-density at which the root-mean-square momentum of the positron distribution is within 1\% of the thermal value $k_{\rm th}\sim 0.0526$ for a room-temperature gas (of temperature $T=293$ K).
The figure shows that $\bar{S}$ reaches its steady state value close to the shoulder time, impressively demonstrating how the spectra can be used to probe positron cooling times.

Figure~\ref{fig:sw1} shows $\bar{W}(\tau)$ and $\bar{S}(\tau)$ (obtained with $\epsilon_W=2.0$ and $\epsilon_S=1.0$ keV) calculated using the detector-convolved spectra excluding and including depletion of the positron distribution due to annihilation, for positrons initially distributed either uniformly in energy or with energy equal to the Ps-formation threshold \footnote{The initial distribution of constant energy is unphysical, but provides an upper-limit on the cooling times. See \cite{DGG_cool} for more details on the choice of the initial distribution.}.
Also marked are the experimental shoulder lengths $\tau_{\rm s}$ \cite{Coleman:1975,Wright:1985}, and calculated $\tau_{\rm s}$ and complete thermalization times from \cite{DGG_cool}.
In general, as the positrons cool, the annihilation spectrum becomes narrower, so the $\bar{S}$ parameter increases and the $\bar{W}$ parameter decreases with time, before reaching a steady-state value at thermalization.
For lighter atoms (He and Ne) the initial positron energy distribution (i.e., uniform vs monoenergetic) does not play much role. Also, for these atoms the annihilation rate remains much smaller than the cooling rates at all times. As a result, depletion of positrons during the cooling process does not affect the time dependence of the $\gamma$ spectra. 
For Ne, the fraction of positrons that survive beyond $\tau\sim 8000$ ns amg is practically zero (leading to poor statistics), owing to positrons becoming `trapped' in the deep momentum-transfer minimum, where cooling is slow \cite{DGG_cool}. Given the insensitivity to the initial distribution, accurate measurements of $\bar{W}$ and $\bar{S}$ could confirm the theoretical predictions.
In contrast, for Ar, Kr and Xe, the initial positron distribution has a sizeable effect on the time evolution of the spectra. The more physical uniform-energy distribution leads to fast evolution of $\bar{S}$ and $\bar{W}$ at earlier times towards the final `thermalized' values. For these atoms, the information provided by AMOC measurements, combined with theoretical studies, could enable determination of the form of the initial distributions, about which little is currently known. 
On the other hand, the effect of positron depletion slows down the time evolution of $\bar{S}$ and $\bar{W}$, which is particularly noticeable in Xe. In this system the positron annihilation rate competes strongly with the positron cooling. Due to the strong peaking of $Z_{\rm eff}(k)$ at small $k$, annihilation effectively removes the slowest positrons, impeding thermalization [and in fact, leading to a non-Maxwell-Boltzmann asymptotic momentum distribution $f_{\infty}(k)]$ \cite{DGG_cool}.

\emph{Summary.---}Many-body theory based calculations of time-varying $\gamma$ spectra for positron annihilation on noble gases have been presented. The benchmark results demonstrate that the spectra provide a sensitive probe of positron cooling, which is complementary to positron lifetime spectroscopy. 

\begin{acknowledgments}
\emph{Acknowledgements.---}{DGG thanks Gleb Gribakin for insightful discussions and useful comments on the manuscript.} 
DGG is supported by the UK Engineering and Physical Sciences Research Council, grant EP/N007948/1.
\end{acknowledgments}

\appendix

\begin{thebibliography}{43}%
\makeatletter
\providecommand \@ifxundefined [1]{%
 \@ifx{#1\undefined}
}%
\providecommand \@ifnum [1]{%
 \ifnum #1\expandafter \@firstoftwo
 \else \expandafter \@secondoftwo
 \fi
}%
\providecommand \@ifx [1]{%
 \ifx #1\expandafter \@firstoftwo
 \else \expandafter \@secondoftwo
 \fi
}%
\providecommand \natexlab [1]{#1}%
\providecommand \enquote  [1]{``#1''}%
\providecommand \bibnamefont  [1]{#1}%
\providecommand \bibfnamefont [1]{#1}%
\providecommand \citenamefont [1]{#1}%
\providecommand \href@noop [0]{\@secondoftwo}%
\providecommand \href [0]{\begingroup \@sanitize@url \@href}%
\providecommand \@href[1]{\@@startlink{#1}\@@href}%
\providecommand \@@href[1]{\endgroup#1\@@endlink}%
\providecommand \@sanitize@url [0]{\catcode `\\12\catcode `\$12\catcode
  `\&12\catcode `\#12\catcode `\^12\catcode `\_12\catcode `\%12\relax}%
\providecommand \@@startlink[1]{}%
\providecommand \@@endlink[0]{}%
\providecommand \url  [0]{\begingroup\@sanitize@url \@url }%
\providecommand \@url [1]{\endgroup\@href {#1}{\urlprefix }}%
\providecommand \urlprefix  [0]{URL }%
\providecommand \Eprint [0]{\href }%
\providecommand \doibase [0]{http://dx.doi.org/}%
\providecommand \selectlanguage [0]{\@gobble}%
\providecommand \bibinfo  [0]{\@secondoftwo}%
\providecommand \bibfield  [0]{\@secondoftwo}%
\providecommand \translation [1]{[#1]}%
\providecommand \BibitemOpen [0]{}%
\providecommand \bibitemStop [0]{}%
\providecommand \bibitemNoStop [0]{.\EOS\space}%
\providecommand \EOS [0]{\spacefactor3000\relax}%
\providecommand \BibitemShut  [1]{\csname bibitem#1\endcsname}%
\let\auto@bib@innerbib\@empty
\bibitem [{\citenamefont {Green}\ and\ \citenamefont
  {Gribakin}(2015)}]{DGG:2015:core}%
  \BibitemOpen
  \bibfield  {author} {\bibinfo {author} {\bibfnamefont {D.~G.}\ \bibnamefont
  {Green}}\ and\ \bibinfo {author} {\bibfnamefont {G.~F.}\ \bibnamefont
  {Gribakin}},\ }\href {\doibase 10.1103/PhysRevLett.114.093201} {\bibfield
  {journal} {\bibinfo  {journal} {Phys. Rev. Lett.}\ }\textbf {\bibinfo
  {volume} {114}},\ \bibinfo {pages} {093201} (\bibinfo {year}
  {2015})}\BibitemShut {NoStop}%
\bibitem [{\citenamefont {Schultz}\ and\ \citenamefont
  {Lynn}(1988)}]{RevModPhys.60.701}%
  \BibitemOpen
  \bibfield  {author} {\bibinfo {author} {\bibfnamefont {P.~J.}\ \bibnamefont
  {Schultz}}\ and\ \bibinfo {author} {\bibfnamefont {K.~G.}\ \bibnamefont
  {Lynn}},\ }\href {\doibase 10.1103/RevModPhys.60.701} {\bibfield  {journal}
  {\bibinfo  {journal} {Rev. Mod. Phys.}\ }\textbf {\bibinfo {volume} {60}},\
  \bibinfo {pages} {701} (\bibinfo {year} {1988})}\BibitemShut {NoStop}%
\bibitem [{\citenamefont {Puska}\ and\ \citenamefont
  {Nieminen}(1994)}]{RevModPhys.66.841}%
  \BibitemOpen
  \bibfield  {author} {\bibinfo {author} {\bibfnamefont {M.~J.}\ \bibnamefont
  {Puska}}\ and\ \bibinfo {author} {\bibfnamefont {R.~M.}\ \bibnamefont
  {Nieminen}},\ }\href {\doibase 10.1103/RevModPhys.66.841} {\bibfield
  {journal} {\bibinfo  {journal} {Rev. Mod. Phys.}\ }\textbf {\bibinfo {volume}
  {66}},\ \bibinfo {pages} {841} (\bibinfo {year} {1994})}\BibitemShut
  {NoStop}%
\bibitem [{\citenamefont {Tuomisto}\ and\ \citenamefont
  {Makkonen}(2013)}]{RMPpossolids2013}%
  \BibitemOpen
  \bibfield  {author} {\bibinfo {author} {\bibfnamefont {F.}~\bibnamefont
  {Tuomisto}}\ and\ \bibinfo {author} {\bibfnamefont {I.}~\bibnamefont
  {Makkonen}},\ }\href {\doibase 10.1103/RevModPhys.85.1583} {\bibfield
  {journal} {\bibinfo  {journal} {Rev. Mod. Phys.}\ }\textbf {\bibinfo {volume}
  {85}},\ \bibinfo {pages} {1583} (\bibinfo {year} {2013})}\BibitemShut
  {NoStop}%
\bibitem [{\citenamefont {Hugenschmidt}(2016)}]{hugreview}%
  \BibitemOpen
  \bibfield  {author} {\bibinfo {author} {\bibfnamefont {C.}~\bibnamefont
  {Hugenschmidt}},\ }\href {\doibase 10.1016/j.surfrep.2016.09.002} {\bibfield  {journal}
  {\bibinfo  {journal} {Surf. Sci. Rep.}\ }\textbf {\bibinfo {volume} {71}},\
  \bibinfo {pages} {547 } (\bibinfo {year} {2016})}\BibitemShut {NoStop}%
\bibitem [{\citenamefont {Stoll}\ \emph {et~al.}(1991)\citenamefont {Stoll},
  \citenamefont {M.~Koch},\ and\ \citenamefont {Major}}]{amoc}%
  \BibitemOpen
  \bibfield  {author} {\bibinfo {author} {\bibfnamefont {H.}~\bibnamefont
  {Stoll}}, \bibinfo {author} {\bibfnamefont {K.~M.}\ \bibnamefont {M.~Koch}},
  \ and\ \bibinfo {author} {\bibfnamefont {J.}~\bibnamefont {Major}},\
  }\href@noop {} {\bibfield  {journal} {\bibinfo  {journal} {Nucl. Instrum. and
  Meth. B}\ }\textbf {\bibinfo {volume} {582}},\ \bibinfo {pages} {56}
  (\bibinfo {year} {1991})}\BibitemShut {NoStop}%
\bibitem [{\citenamefont {Coleman}(2000)}]{posbeams}%
  \BibitemOpen
  \bibinfo {editor} {\bibfnamefont {P.}~\bibnamefont {Coleman}},\ ed.,\
  \href@noop {} {\emph {\bibinfo {title} {Positron beams and their
  applications}}}\ (\bibinfo  {publisher} {World Scientific},\ \bibinfo {year}
  {2000})\BibitemShut {NoStop}%
\bibitem [{\citenamefont {Siegle}\ \emph {et~al.}(1997)\citenamefont {Siegle},
  \citenamefont {Stoll}, \citenamefont {Castellaz}, \citenamefont {Major},
  \citenamefont {Schneider},\ and\ \citenamefont {Seeger}}]{AMOC:1997}%
  \BibitemOpen
  \bibfield  {author} {\bibinfo {author} {\bibfnamefont {A.}~\bibnamefont
  {Siegle}}, \bibinfo {author} {\bibfnamefont {H.}~\bibnamefont {Stoll}},
  \bibinfo {author} {\bibfnamefont {P.}~\bibnamefont {Castellaz}}, \bibinfo
  {author} {\bibfnamefont {J.}~\bibnamefont {Major}}, \bibinfo {author}
  {\bibfnamefont {H.}~\bibnamefont {Schneider}}, \ and\ \bibinfo {author}
  {\bibfnamefont {A.}~\bibnamefont {Seeger}},\ }\href {\doibase 10.1016/S0169-4332(96)01043-4} {\bibfield  {journal}
  {\bibinfo  {journal} {App. Surf. Sci.}\ }\textbf {\bibinfo {volume} {116}},\
  \bibinfo {pages} {140 } (\bibinfo {year} {1997})}\BibitemShut {NoStop}%
\bibitem [{\citenamefont {Engbrecht}(2004)}]{Engbrecht:AMOC}%
  \BibitemOpen
  \bibfield  {author} {\bibinfo {author} {\bibfnamefont {J.}~\bibnamefont
  {Engbrecht}},\ }\href {\doibase 10.1016/j.nimb.2004.03.041}
  {\bibfield  {journal} {\bibinfo  {journal} {Nuc. Instrum. Meth. B}\ }\textbf
  {\bibinfo {volume} {221}},\ \bibinfo {pages} {119 } (\bibinfo {year}
  {2004})}\BibitemShut {NoStop}%
\bibitem [{\citenamefont {Ackermann}\ \emph {et~al.}(2016)\citenamefont
  {Ackermann}, \citenamefont {L{\"o}we}, \citenamefont {Dickmann},
  \citenamefont {Mitteneder}, \citenamefont {Sperr}, \citenamefont {Egger},
  \citenamefont {Reiner},\ and\ \citenamefont {Dollinger}}]{AMOC:2016}%
  \BibitemOpen
  \bibfield  {author} {\bibinfo {author} {\bibfnamefont {U.}~\bibnamefont
  {Ackermann}}, \bibinfo {author} {\bibfnamefont {B.}~\bibnamefont {L{\"o}we}},
  \bibinfo {author} {\bibfnamefont {M.}~\bibnamefont {Dickmann}}, \bibinfo
  {author} {\bibfnamefont {J.}~\bibnamefont {Mitteneder}}, \bibinfo {author}
  {\bibfnamefont {P.}~\bibnamefont {Sperr}}, \bibinfo {author} {\bibfnamefont
  {W.}~\bibnamefont {Egger}}, \bibinfo {author} {\bibfnamefont
  {M.}~\bibnamefont {Reiner}}, \ and\ \bibinfo {author} {\bibfnamefont
  {G.}~\bibnamefont {Dollinger}},\ }\href
  {http://stacks.iop.org/1367-2630/18/i=11/a=113030} {\bibfield  {journal}
  {\bibinfo  {journal} {New J. Phys.}\ }\textbf {\bibinfo {volume} {18}},\
  \bibinfo {pages} {113030} (\bibinfo {year} {2016})}\BibitemShut {NoStop}%
\bibitem [{\citenamefont {Engbrecht}\ \emph {et~al.}(2008)\citenamefont
  {Engbrecht}, \citenamefont {Erickson}, \citenamefont {Johnson}, \citenamefont
  {Kolan}, \citenamefont {Legard}, \citenamefont {Lund}, \citenamefont
  {Nyflot},\ and\ \citenamefont {Paulsen}}]{AMOC:psmt}%
  \BibitemOpen
  \bibfield  {author} {\bibinfo {author} {\bibfnamefont {J.~J.}\ \bibnamefont
  {Engbrecht}}, \bibinfo {author} {\bibfnamefont {M.~J.}\ \bibnamefont
  {Erickson}}, \bibinfo {author} {\bibfnamefont {C.~P.}\ \bibnamefont
  {Johnson}}, \bibinfo {author} {\bibfnamefont {A.~J.}\ \bibnamefont {Kolan}},
  \bibinfo {author} {\bibfnamefont {A.~E.}\ \bibnamefont {Legard}}, \bibinfo
  {author} {\bibfnamefont {S.~P.}\ \bibnamefont {Lund}}, \bibinfo {author}
  {\bibfnamefont {M.~J.}\ \bibnamefont {Nyflot}}, \ and\ \bibinfo {author}
  {\bibfnamefont {J.~D.}\ \bibnamefont {Paulsen}},\ }\href {\doibase 10.1103/PhysRevA.77.012711} {\bibfield  {journal} {\bibinfo  {journal} {Phys.
  Rev. A}\ }\textbf {\bibinfo {volume} {77}},\ \bibinfo {pages} {012711}
  (\bibinfo {year} {2008})}\BibitemShut {NoStop}%
\bibitem [{\citenamefont {Sano}\ \emph {et~al.}(2014)\citenamefont {Sano},
  \citenamefont {Kino}, \citenamefont {Oka},\ and\ \citenamefont
  {Sekine}}]{AMOC:ar}%
  \BibitemOpen
  \bibfield  {author} {\bibinfo {author} {\bibfnamefont {Y.}~\bibnamefont
  {Sano}}, \bibinfo {author} {\bibfnamefont {Y.}~\bibnamefont {Kino}}, \bibinfo
  {author} {\bibfnamefont {T.}~\bibnamefont {Oka}}, \ and\ \bibinfo {author}
  {\bibfnamefont {T.}~\bibnamefont {Sekine}},\ }\href {\doibase 10.7567/JJAPCP.2.011004} {\bibfield  {journal} {\bibinfo  {journal} {Jpn. J.
  Appl. Phys. Conf. Proc.}\ }\textbf {\bibinfo {volume} {2}},\ \bibinfo {pages}
  {011004} (\bibinfo {year} {2014})}\BibitemShut {NoStop}%
\bibitem [{\citenamefont {Al-Qaradawi}\ \emph {et~al.}(2000)\citenamefont
  {Al-Qaradawi}, \citenamefont {Charlton}, \citenamefont {Borozan},
  \citenamefont {Whitehead},\ and\ \citenamefont {Borozan}}]{AlQaradawi:2000}%
  \BibitemOpen
  \bibfield  {author} {\bibinfo {author} {\bibfnamefont {I.}~\bibnamefont
  {Al-Qaradawi}}, \bibinfo {author} {\bibfnamefont {M.}~\bibnamefont
  {Charlton}}, \bibinfo {author} {\bibfnamefont {I.}~\bibnamefont {Borozan}},
  \bibinfo {author} {\bibfnamefont {R.}~\bibnamefont {Whitehead}}, \ and\
  \bibinfo {author} {\bibfnamefont {I.}~\bibnamefont {Borozan}},\ }\href
  {http://stacks.iop.org/0953-4075/33/i=14/a=309} {\bibfield  {journal}
  {\bibinfo  {journal} {J. Phys. B}\ }\textbf {\bibinfo {volume} {33}},\
  \bibinfo {pages} {2725} (\bibinfo {year} {2000})}\BibitemShut {NoStop}%
\bibitem [{\citenamefont {Natisin}\ \emph {et~al.}(2014)\citenamefont
  {Natisin}, \citenamefont {Danielson},\ and\ \citenamefont
  {Surko}}]{Natisin:2014}%
  \BibitemOpen
  \bibfield  {author} {\bibinfo {author} {\bibfnamefont {M.~R.}\ \bibnamefont
  {Natisin}}, \bibinfo {author} {\bibfnamefont {J.~R.}\ \bibnamefont
  {Danielson}}, \ and\ \bibinfo {author} {\bibfnamefont {C.~M.}\ \bibnamefont
  {Surko}},\ }\href {http://stacks.iop.org/0953-4075/47/i=22/a=225209}
  {\bibfield  {journal} {\bibinfo  {journal} {J. Phys. B}\ }\textbf {\bibinfo
  {volume} {47}},\ \bibinfo {pages} {225209} (\bibinfo {year}
  {2014})}\BibitemShut {NoStop}%
\bibitem [{\citenamefont {Natisin}\ \emph {et~al.}(2016)\citenamefont
  {Natisin}, \citenamefont {Danielson},\ and\ \citenamefont
  {Surko}}]{Natisin:2016}%
  \BibitemOpen
  \bibfield  {author} {\bibinfo {author} {\bibfnamefont {M.~R.}\ \bibnamefont
  {Natisin}}, \bibinfo {author} {\bibfnamefont {J.~R.}\ \bibnamefont
  {Danielson}}, \ and\ \bibinfo {author} {\bibfnamefont {C.~M.}\ \bibnamefont
  {Surko}},\ }\href {\doibase 10.1063/1.4939854} {\bibfield  {journal}
  {\bibinfo  {journal} {Appl. Phys. Lett.}\ }\textbf {\bibinfo {volume}
  {108}},\ \bibinfo {pages} {024102} (\bibinfo {year} {2016})}\BibitemShut
  {NoStop}%
\bibitem [{\citenamefont {Griffith}\ and\ \citenamefont
  {Heyland}(1978)}]{Griffith:1979}%
  \BibitemOpen
  \bibfield  {author} {\bibinfo {author} {\bibfnamefont {T.~C.}\ \bibnamefont
  {Griffith}}\ and\ \bibinfo {author} {\bibfnamefont {G.~R.}\ \bibnamefont
  {Heyland}},\ }\href {\doibase 10.1016/0370-1573(78)90127-8}
  {\bibfield  {journal} {\bibinfo  {journal} {Phys.~Rep.}\ }\textbf {\bibinfo
  {volume} {39}},\ \bibinfo {pages} {169 } (\bibinfo {year}
  {1978})}\BibitemShut {NoStop}%
\bibitem [{\citenamefont {Charlton}(1985)}]{Charlton:1985}%
  \BibitemOpen
  \bibfield  {author} {\bibinfo {author} {\bibfnamefont {M.}~\bibnamefont
  {Charlton}},\ }\href {http://stacks.iop.org/0034-4885/48/i=6/a=001}
  {\bibfield  {journal} {\bibinfo  {journal} {Rep. Prog. Phys.}\ }\textbf
  {\bibinfo {volume} {48}},\ \bibinfo {pages} {737} (\bibinfo {year}
  {1985})}\BibitemShut {NoStop}%
\bibitem [{\citenamefont {Green}()}]{DGG_cool}%
  \BibitemOpen
  \bibfield  {author} {\bibinfo {author} {\bibfnamefont {D.~G.}\ \bibnamefont
  {Green}},\ }\href@noop {} {\enquote {\bibinfo {title} {Positron cooling and
  annihilation in noble gases},}\ }\Eprint
  {http://arxiv.org/abs/arXiv:1706.01434v1} {arXiv:1706.01434v1 (2017)} \BibitemShut
  {NoStop}%
\bibitem [{\citenamefont {Dzuba}\ \emph {et~al.}(1993)\citenamefont {Dzuba},
  \citenamefont {Flambaum}, \citenamefont {King}, \citenamefont {Miller},\ and\
  \citenamefont {Sushkov}}]{PhysScripta.46.248}%
  \BibitemOpen
  \bibfield  {author} {\bibinfo {author} {\bibfnamefont {V.~A.}\ \bibnamefont
  {Dzuba}}, \bibinfo {author} {\bibfnamefont {V.~V.}\ \bibnamefont {Flambaum}},
  \bibinfo {author} {\bibfnamefont {W.~A.}\ \bibnamefont {King}}, \bibinfo
  {author} {\bibfnamefont {B.~N.}\ \bibnamefont {Miller}}, \ and\ \bibinfo
  {author} {\bibfnamefont {O.~P.}\ \bibnamefont {Sushkov}},\ }\href {\doibase 10.1088/0031-8949/1993/T46/039} {\bibfield  {journal} {\bibinfo  {journal}
  {Phys. Scripta}\ }\textbf {\bibinfo {volume} {T46}},\ \bibinfo {pages} {248}
  (\bibinfo {year} {1993})}\BibitemShut {NoStop}%
\bibitem [{\citenamefont {Dzuba}\ \emph {et~al.}(1996)\citenamefont {Dzuba},
  \citenamefont {Flambaum}, \citenamefont {Gribakin},\ and\ \citenamefont
  {King}}]{dzuba_mbt_noblegas}%
  \BibitemOpen
  \bibfield  {author} {\bibinfo {author} {\bibfnamefont {V.~A.}\ \bibnamefont
  {Dzuba}}, \bibinfo {author} {\bibfnamefont {V.~V.}\ \bibnamefont {Flambaum}},
  \bibinfo {author} {\bibfnamefont {G.~F.}\ \bibnamefont {Gribakin}}, \ and\
  \bibinfo {author} {\bibfnamefont {W.~A.}\ \bibnamefont {King}},\ }\href
  {\doibase 10.1088/0953-4075/29/14/024} {\bibfield  {journal} {\bibinfo
  {journal} {J. Phys. B}\ }\textbf {\bibinfo {volume} {29}},\ \bibinfo {pages}
  {3151} (\bibinfo {year} {1996})}\BibitemShut {NoStop}%
\bibitem [{\citenamefont {Green}\ \emph {et~al.}(2014)\citenamefont {Green},
  \citenamefont {Ludlow},\ and\ \citenamefont {Gribakin}}]{DGG_posnobles}%
  \BibitemOpen
  \bibfield  {author} {\bibinfo {author} {\bibfnamefont {D.~G.}\ \bibnamefont
  {Green}}, \bibinfo {author} {\bibfnamefont {J.~A.}\ \bibnamefont {Ludlow}}, \
  and\ \bibinfo {author} {\bibfnamefont {G.~F.}\ \bibnamefont {Gribakin}},\
  }\href {\doibase 10.1103/PhysRevA.90.032712} {\bibfield  {journal} {\bibinfo
  {journal} {Phys. Rev. A}\ }\textbf {\bibinfo {volume} {90}},\ \bibinfo
  {pages} {032712} (\bibinfo {year} {2014})}\BibitemShut {NoStop}%
\bibitem [{\citenamefont {Berestetskii}\ \emph {et~al.}(1982)\citenamefont
  {Berestetskii}, \citenamefont {Lifshitz},\ and\ \citenamefont
  {Pitaevskii}}]{qed}%
  \BibitemOpen
  \bibfield  {author} {\bibinfo {author} {\bibfnamefont {V.~B.}\ \bibnamefont
  {Berestetskii}}, \bibinfo {author} {\bibfnamefont {E.~M.}\ \bibnamefont
  {Lifshitz}}, \ and\ \bibinfo {author} {\bibfnamefont {L.~P.}\ \bibnamefont
  {Pitaevskii}},\ }\href@noop {} {\emph {\bibinfo {title} {Quantum
  Electrodynamics}}},\ \bibinfo {edition} {2nd}\ ed.\ (\bibinfo  {publisher}
  {Pergamon, Oxford},\ \bibinfo {year} {1982})\BibitemShut {NoStop}%
\bibitem [{\citenamefont {Dunlop}\ and\ \citenamefont
  {Gribakin}(2006)}]{Dunlop:gamma}%
  \BibitemOpen
  \bibfield  {author} {\bibinfo {author} {\bibfnamefont {L.~J.~M.}\
  \bibnamefont {Dunlop}}\ and\ \bibinfo {author} {\bibfnamefont {G.~F.}\
  \bibnamefont {Gribakin}},\ }\href {\doibase 10.1088/0953-4075/39/7/008}
  {\bibfield  {journal} {\bibinfo  {journal} {J. Phys. B}\ }\textbf {\bibinfo
  {volume} {39}},\ \bibinfo {pages} {1647} (\bibinfo {year}
  {2006})}\BibitemShut {NoStop}%
\bibitem [{\citenamefont {Green}\ and\ \citenamefont
  {Gribakin}()}]{DGG_corelong}%
  \BibitemOpen
  \bibfield  {author} {\bibinfo {author} {\bibfnamefont {D.~G.}\ \bibnamefont
  {Green}}\ and\ \bibinfo {author} {\bibfnamefont {G.~F.}\ \bibnamefont
  {Gribakin}},\ }\href@noop {} {\enquote {\bibinfo {title} {Positron
  annihilation on core and valence electrons},}\ }\Eprint
  {http://arxiv.org/abs/arXiv:1502.08045} {arXiv:1502.08045} \BibitemShut
  {NoStop}%
\bibitem [{\citenamefont {Fraser}(1968)}]{Fraser}%
  \BibitemOpen
  \bibfield  {author} {\bibinfo {author} {\bibfnamefont {P.~A.}\ \bibnamefont
  {Fraser}},\ }\href@noop {} {\bibfield  {journal} {\bibinfo  {journal} {Adv.
  At. Mol. Phys.}\ }\textbf {\bibinfo {volume} {4}},\ \bibinfo {pages} {63}
  (\bibinfo {year} {1968})}\BibitemShut {NoStop}%
\bibitem [{\citenamefont {Pomeranchuk}(1949)}]{Pomeranchuk}%
  \BibitemOpen
  \bibfield  {author} {\bibinfo {author} {\bibfnamefont {I.}~\bibnamefont
  {Pomeranchuk}},\ }\href@noop {} {\bibfield  {journal} {\bibinfo  {journal}
  {Zh. Eksp. Teor. Fiz.}\ }\textbf {\bibinfo {volume} {19}},\ \bibinfo {pages}
  {183} (\bibinfo {year} {1949})}\BibitemShut {NoStop}%
\bibitem [{Note1()}]{Note1}%
  \BibitemOpen
  \bibinfo {note} {After integrating over the directions of ${\protect \bf P}$,
  all positron partial waves contribute to the amplitude incoherently, such
  that it can be calculated independently for each positron angular momentum
  $\ell _{\varepsilon }$.}\BibitemShut {Stop}%
\bibitem [{\citenamefont {Green}\ and\ \citenamefont
  {Gribakin}(2013)}]{DGG_hlike}%
  \BibitemOpen
  \bibfield  {author} {\bibinfo {author} {\bibfnamefont {D.~G.}\ \bibnamefont
  {Green}}\ and\ \bibinfo {author} {\bibfnamefont {G.~F.}\ \bibnamefont
  {Gribakin}},\ }\href {\doibase 10.1103/PhysRevA.88.032708} {\bibfield
  {journal} {\bibinfo  {journal} {Phys. Rev. A}\ }\textbf {\bibinfo {volume}
  {88}},\ \bibinfo {pages} {032708} (\bibinfo {year} {2013})}\BibitemShut
  {NoStop}%
\bibitem [{Note2()}]{Note2}%
  \BibitemOpen
  \bibinfo {note} {At low momenta the Wigner threshold law predicts the partial
  rate $Z_{\protect \rm eff}^{(\ell _{\varepsilon })}(k)\sim k^{2\ell
  _{\varepsilon }}$ \cite {quantummechanics} and thus the $s$-wave contribution
  dominates the spectra. However, the contribution of $p$ and $d$-waves
  increases and become comparable at positron momenta close to the
  positronium-formation threshold.}\BibitemShut {Stop}%
\bibitem [{Note3()}]{Note3}%
  \BibitemOpen
  \bibinfo {note} {The $\gamma $ spectra for thermal positrons presented in
  \cite {DGG:2015:core} were found to be in excellent agreement with experiment
  \cite {PhysRevLett.79.39}. A slight discrepancy was, however, observed in the
  wings of the spectrum for Xe, where the MBT underestimated the experimental
  result. It was proffered that this may be remedied by considering the
  thermally averaged spectrum. It has been calculated in the present work for
  all the noble gases, but however, has been found to be almost
  indistinguishable from the $\gamma $ spectra calculated using positrons with
  $k=0.04$ a.u. It remains to investigate whether the discrepancy is thus due
  to the neglect of relativistic effects that may enhance the core annihilation
  spectra. Relativistic MBT calculations are required to confirm
  this.}\BibitemShut {Stop}%
\bibitem [{\citenamefont {Goldanski}\ and\ \citenamefont
  {Sayasov}(1968)}]{Goldanski}%
  \BibitemOpen
  \bibfield  {author} {\bibinfo {author} {\bibfnamefont {V.~I.}\ \bibnamefont
  {Goldanski}}\ and\ \bibinfo {author} {\bibfnamefont {Y.~S.}\ \bibnamefont
  {Sayasov}},\ }\href@noop {} {\bibfield  {journal} {\bibinfo  {journal} {Phys.
  Lett.}\ }\textbf {\bibinfo {volume} {13}},\ \bibinfo {pages} {300} (\bibinfo
  {year} {1968})}\BibitemShut {NoStop}%
\bibitem [{\citenamefont {Iwata}\ \emph {et~al.}(1997)\citenamefont {Iwata},
  \citenamefont {Gribakin}, \citenamefont {Greaves},\ and\ \citenamefont
  {Surko}}]{PhysRevLett.79.39}%
  \BibitemOpen
  \bibfield  {author} {\bibinfo {author} {\bibfnamefont {K.}~\bibnamefont
  {Iwata}}, \bibinfo {author} {\bibfnamefont {G.~F.}\ \bibnamefont {Gribakin}},
  \bibinfo {author} {\bibfnamefont {R.~G.}\ \bibnamefont {Greaves}}, \ and\
  \bibinfo {author} {\bibfnamefont {C.~M.}\ \bibnamefont {Surko}},\ }\href
  {\doibase 10.1103/PhysRevLett.79.39} {\bibfield  {journal} {\bibinfo
  {journal} {Phys. Rev. Lett.}\ }\textbf {\bibinfo {volume} {79}},\ \bibinfo
  {pages} {39} (\bibinfo {year} {1997})}\BibitemShut {NoStop}%
\bibitem [{Note4()}]{Note4}%
  \BibitemOpen
  \bibinfo {note} {This holds for a spectrum convolved with a
  detector-resolution function $D(\epsilon )$ normalized as $\DOTSI \intop
  \ilimits@ _{-\infty }^{\infty } D(\epsilon )d\epsilon =1$.}\BibitemShut
  {Stop}%
\bibitem [{\citenamefont {Falk}\ and\ \citenamefont {Jones}(1964)}]{Falk:1964}%
  \BibitemOpen
  \bibfield  {author} {\bibinfo {author} {\bibfnamefont {W.~R.}\ \bibnamefont
  {Falk}}\ and\ \bibinfo {author} {\bibfnamefont {G.}~\bibnamefont {Jones}},\
  }\href {\doibase 10.1139/p64-160} {\bibfield  {journal} {\bibinfo  {journal}
  {Can. J. Phys.}\ }\textbf {\bibinfo {volume} {42}},\ \bibinfo {pages} {1751}
  (\bibinfo {year} {1964})}\BibitemShut {NoStop}%
\bibitem [{\citenamefont {Tao}\ \emph {et~al.}(1964)\citenamefont {Tao},
  \citenamefont {Bell},\ and\ \citenamefont {Green}}]{Tao:1964}%
  \BibitemOpen
  \bibfield  {author} {\bibinfo {author} {\bibfnamefont {S.~J.}\ \bibnamefont
  {Tao}}, \bibinfo {author} {\bibfnamefont {J.}~\bibnamefont {Bell}}, \ and\
  \bibinfo {author} {\bibfnamefont {J.~H.}\ \bibnamefont {Green}},\ }\href
  {http://stacks.iop.org/0370-1328/83/i=3/a=312} {\bibfield  {journal}
  {\bibinfo  {journal} {Proc.~Phys.~Soc.}\ }\textbf {\bibinfo {volume} {83}},\
  \bibinfo {pages} {453} (\bibinfo {year} {1964})}\BibitemShut {NoStop}%
\bibitem [{\citenamefont {Paul}(1964)}]{Paul:1964}%
  \BibitemOpen
  \bibfield  {author} {\bibinfo {author} {\bibfnamefont {D.~A.~L.}\
  \bibnamefont {Paul}},\ }\href {http://stacks.iop.org/0370-1328/84/i=4/a=312}
  {\bibfield  {journal} {\bibinfo  {journal} {Proc.~Phys.~Soc.}\ }\textbf
  {\bibinfo {volume} {84}},\ \bibinfo {pages} {563} (\bibinfo {year}
  {1964})}\BibitemShut {NoStop}%
\bibitem [{\citenamefont {Farazdel}\ and\ \citenamefont
  {Epstein}(1977)}]{Farazdel:1977}%
  \BibitemOpen
  \bibfield  {author} {\bibinfo {author} {\bibfnamefont {A.}~\bibnamefont
  {Farazdel}}\ and\ \bibinfo {author} {\bibfnamefont {I.~R.}\ \bibnamefont
  {Epstein}},\ }\href {\doibase 10.1103/PhysRevA.16.518} {\bibfield  {journal}
  {\bibinfo  {journal} {Phys. Rev. A}\ }\textbf {\bibinfo {volume} {16}},\
  \bibinfo {pages} {518} (\bibinfo {year} {1977})}\BibitemShut {NoStop}%
\bibitem [{Note5()}]{Note5}%
  \BibitemOpen
  \bibinfo {note} {For He the final distribution is the Maxwell-Boltzmann
  distribution. For the heavier gases, most notably Xe, the low-momentum side
  of the final distribution is suppressed relative to the Maxwell Boltzmann
  distribution. This is a result of the the rapid depletion of low-momentum
  positrons due to the vigorous increase in $Z_{\protect \rm eff}$ at low
  momenta.}\BibitemShut {Stop}%
\bibitem [{\citenamefont {Coleman}\ \emph {et~al.}(1975)\citenamefont
  {Coleman}, \citenamefont {Griffith}, \citenamefont {Heyland},\ and\
  \citenamefont {Killeen}}]{Coleman:1975}%
  \BibitemOpen
  \bibfield  {author} {\bibinfo {author} {\bibfnamefont {P.~G.}\ \bibnamefont
  {Coleman}}, \bibinfo {author} {\bibfnamefont {T.~C.}\ \bibnamefont
  {Griffith}}, \bibinfo {author} {\bibfnamefont {G.~R.}\ \bibnamefont
  {Heyland}}, \ and\ \bibinfo {author} {\bibfnamefont {T.~L.}\ \bibnamefont
  {Killeen}},\ }\href {http://stacks.iop.org/0022-3700/8/i=10/a=021} {\bibfield
   {journal} {\bibinfo  {journal} {J. Phys. B}\ }\textbf {\bibinfo {volume}
  {8}},\ \bibinfo {pages} {1734} (\bibinfo {year} {1975})}\BibitemShut
  {NoStop}%
\bibitem [{\citenamefont {Wright}\ \emph {et~al.}(1985)\citenamefont {Wright},
  \citenamefont {Charlton}, \citenamefont {Griffith},\ and\ \citenamefont
  {Heyland}}]{Wright:1985}%
  \BibitemOpen
  \bibfield  {author} {\bibinfo {author} {\bibfnamefont {G.~L.}\ \bibnamefont
  {Wright}}, \bibinfo {author} {\bibfnamefont {M.}~\bibnamefont {Charlton}},
  \bibinfo {author} {\bibfnamefont {T.~C.}\ \bibnamefont {Griffith}}, \ and\
  \bibinfo {author} {\bibfnamefont {G.~R.}\ \bibnamefont {Heyland}},\ }\href
  {http://stacks.iop.org/0022-3700/18/i=21/a=019} {\bibfield  {journal}
  {\bibinfo  {journal} {J. Phys. B}\ }\textbf {\bibinfo {volume} {18}},\
  \bibinfo {pages} {4327} (\bibinfo {year} {1985})}\BibitemShut {NoStop}%
\bibitem [{\citenamefont {Paul}\ and\ \citenamefont {Leung}(1968)}]{Paul:1968}%
  \BibitemOpen
  \bibfield  {author} {\bibinfo {author} {\bibfnamefont {D.~A.~L.}\
  \bibnamefont {Paul}}\ and\ \bibinfo {author} {\bibfnamefont {C.~Y.}\
  \bibnamefont {Leung}},\ }\href {http://dx.doi.org/10.1139/p68-648} {\bibfield
   {journal} {\bibinfo  {journal} {Can. J. Phys.}\ }\textbf {\bibinfo {volume}
  {46}},\ \bibinfo {pages} {2779} (\bibinfo {year} {1968})}\BibitemShut
  {NoStop}%
\bibitem [{Note6()}]{Note6}%
  \BibitemOpen
  \bibinfo {note} {The initial distribution of constant energy is unphysical,
  but provides an upper-limit on the cooling times. See \cite {DGG_cool} for
  more details on the choice of the initial distribution.}\BibitemShut {Stop}%
\bibitem [{\citenamefont {Landau}\ and\ \citenamefont
  {Lifshitz}(1977)}]{quantummechanics}%
  \BibitemOpen
  \bibfield  {author} {\bibinfo {author} {\bibfnamefont {L.~D.}\ \bibnamefont
  {Landau}}\ and\ \bibinfo {author} {\bibfnamefont {E.~M.}\ \bibnamefont
  {Lifshitz}},\ }\href@noop {} {\emph {\bibinfo {title} {Quantum Mechanics
  (Non-relativistic Theory) - Third Edition - Course of Theoretical Physics
  Volume 3}}}\ (\bibinfo  {publisher} {Pergamon, Oxford},\ \bibinfo {year}
  {1977})\BibitemShut {NoStop}%
\end{thebibliography}

%

\end{document}